\documentstyle[10pt,epsfig]{elsart}
\begin{document}
\begin{frontmatter}
\title{Nuclear Matter Expansion Parameters from the Measurement of 
	Differential Multiplicities for $\Lambda$ Production  
	in Central $Au+Au$ Collisions at AGS}
\thanks[doe]{This research was supported by the U.S. Department of 
	Energy under Contract No. DOE/ER/40772-24.}
\author{S. Ahmad},
\author{B.E. Bonner},
\author{S.V. Efremov},
\author{G.S. Mutchler},
\author{E.D. Platner},
\author{H.W. Themann}
\address{Rice University, Houston, TX 77005, USA}
\maketitle
\begin{abstract}
       The double differential multiplicities and rapidity
       distributions for $\Lambda$ hyperon production in central
       $Au+Au$ interactions at AGS in the range of rapidities from
       1.7 to 3.2 and the range of transverse kinetic energies from
       0.0 to $0.7~GeV$ are parametrized in terms of the
       the Blast Wave approximation.
       The longitudinal and transverse radial expansion parameters
       and the mean temperature of $\Lambda$ hyperons
       after the freeze-out of the nuclear matter are presented.
       The predictions of the RQMD model with and without mean field
       potentials are compared to our data. Both variants of RQMD are
       parameterized in terms of the Blast Wave model and the
       results of such
       parameterizations are compared to the experimental ones.
       It is found that inclusion of the mean field potentials in RQMD
       is essential to account for the strong expansion observed in
	 the data.
\end{abstract}
\end{frontmatter}
\section{Introduction}
	The collective nuclear flow has been a subject of extensive theoretical
	and experimental investigations for the last few years. It was pointed
	out that, the
	collective flow directly probes the equation of state of nuclear matter
	and thus can provide one with information about the possible phase
	transition \cite{hung,rischke}. 
	Before a conclusion about the phase transition can be made, 
	the nuclear equation of state should 
	be known for the baryon densities far above normal, but where the 
	matter is still in the hadronic phase. However the theoretical statements 
	for the compressibility of hadronic phase or for  
	the stiffness of the equation of state contradict each other 
	\cite{hung,flow_sorge}. Measurements of collective nuclear flow
	may specify the limits of possible compressibilities 
	of hadronic matter.
	
	One usually distinguishes between longitudinal flow, transverse 
	directed flow, elliptic flow and transverse radial expansion, which are
	interrelated forms of a global picture 
	\cite{flow_sorge,flow_olli}. The directed flow comprises
	a correlation between the impact parameter vector and the directions of
	produced particles. For quasi-central heavy-ion
	interactions this correlation vanishes and one encounters 
	the azimuthal symmetry of flow with respect to the impact parameter:
	the transverse radial expansion superposed on the longitudinal expansion.
	The spectra of particles produced in the heavy-ion interactions 
	represent the convolution of stochastic thermal and collective
	macroscopic motion. The shapes of one-particle multiplicities 
	carry part of the information on the collective velocity profile together with
	information on the stochastic fluctuations of the particle velocities.
	Even though it is impossible to deconvolve the spectra precisely, one 
	can obtain a semi-quantitative picture of nuclear matter expansion with certain 
	assumptions about the flow velocity profile. Studying the spectra of
	$\Lambda$ hyperons has several advantages over studying the spectra of 
	pions and protons. First of all, in central heavy-ion collisions 
	the majority of these particles are produced via multi-step 
	interactions, due to the suppression of the strangeness production in
	nucleon-nucleon collisions. Thus, the $\Lambda$ hyperon spectra
	essentially miss a non-hydrodynamical component, which is present 
	in the spectra of protons. Further, unlike the $\pi$ mesons the
	influence of resonances on the shapes of the $\Lambda$ spectrum is
	negligible. And finally, being heavy, the $\Lambda$ hyperons acquire 
	significant momentum due to the collective component of their velocities. 
	
	In this paper we will show that the double differential multiplicities
	for the $\Lambda$ production in central $Au+Au$ collisions at AGS,
	experiment E891, 
	can be described in terms of an aspherical
	blast wave with a single temperature. We will extract the parameters
	of the blast wave from our data and from the predictions of two versions
	of the RQMD model (Relativistic Quantum Molecular Dynamics \cite{flow_sorge}): 
	the cascade version and the version with the mean fields turned on. 
\section{Experimental method}
\begin{figure}[h]
	\centerline{\hbox{
	\psfig{figure=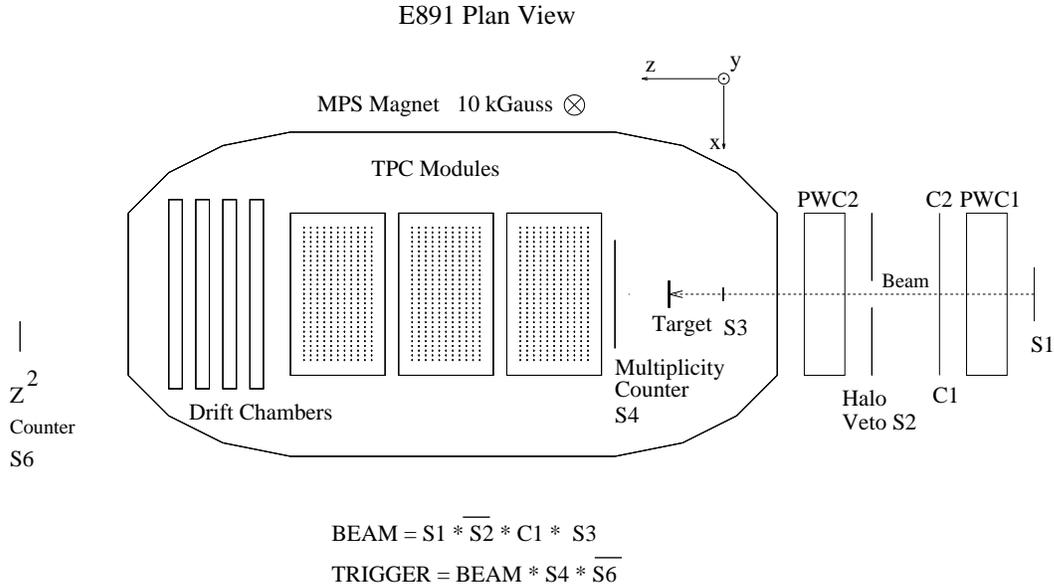,height=3.0in}
	}}
	\caption{The E891 plan view. Dimensions are out of scale.}
	\label{E891}
	\end{figure}

We analyze the experimental data of the E891 collaboration 
on the $\Lambda$ hyperon production in central 
$Au+Au$ collisions at the BNL AGS accelerator \cite{e891}.
The plan view of the E891 apparatus is shown in Figure \ref{E891}. The
$11.6\times A~GeV/c$ $Au$ beam was incident 
on a $0.025~cm~Au$ target (7.5\% radiation lengths). 
The charged particles were tracked in three
Time Projection Chambers (TPC) in the $10~kG$ field of
the Multiple Particle Spectrometer (MPS) magnet. The field volume of the
MPS magnet is 4.60 meters long $\times 1.80$ meters wide $\times 1.20$ 
meters high.
Downstream of the TPC's were four Multiwire Drift Chambers used for track
calibration. The Au target was $80~cm$ upstream of the active volume of the
first TPC. This distance was chosen to maintain adequate two track
separation to reduce track merging in the front TPC module. The beam was
incident on the center of the first TPC. This arrangement covered the
forward hemisphere in the $NN$ center of mass system for rapidities $>1.7$.
The beam intensity varied from 500 to 2000 $Au$ ions per second.

\begin{figure}[h]
	\centerline{\hbox{
	\psfig{figure=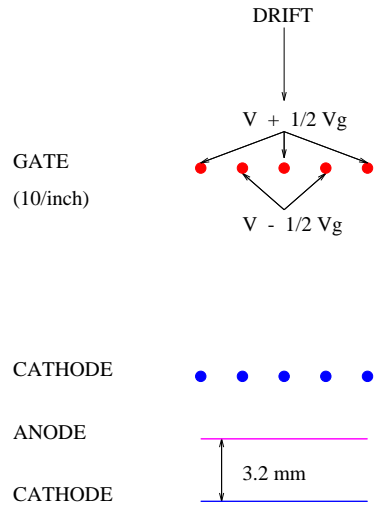,height=3.0in}
	}}
	\caption{Schematic arrangement of electrodes in the detector end cap.}
	\label{ENDCAP}
	\end{figure}

The drift volume of each TPC module is $60~cm$ high $\times 65~cm$ wide 
$\times 47~cm$
long. The drift volume was filled with a low-diffusion gas mixture (79\%
Argon, 16\% Isobutane and 5\% Dimethoxymethane) known to be stable at high
gain. The TPC's were operated at $20~kV$. The high voltage was applied
parallel to the magnetic field.  The three dimensional coordinates of the
hits in the drift volume of each TPC were measured by 12 rows (spaced 3.8
cm apart in the  $z$-direction) of $1~cm$ long anode wires with 256 wires in
each row, ($2.54~mm$ spacing in the $x$ direction). The drift time provides the
vertical coordinate, $y$, measured in 1024 time bins and the wire number
gives the two horizontal coordinates. The $z$ axis points along the beam, the
$x$ axis is horizontal and points towards the bend direction of positive
particles, and the $y$ axis points up with the magnetic field directed along
the $y$ axis. Electrons produced by the charged particles in the drift region
must pass through a gating grid and a wire cathode in order to reach the
anode wires. A second, solid copper cathode, below the anodes helps to
achieve the high gain (Figure \ref{ENDCAP}). The use of short anode wires
precludes making $dE/dx$ measurements. Further details about the TPC
construction are given in \cite{a}.

	\begin{figure}[h]
	\centerline{\hbox{
	\psfig{figure=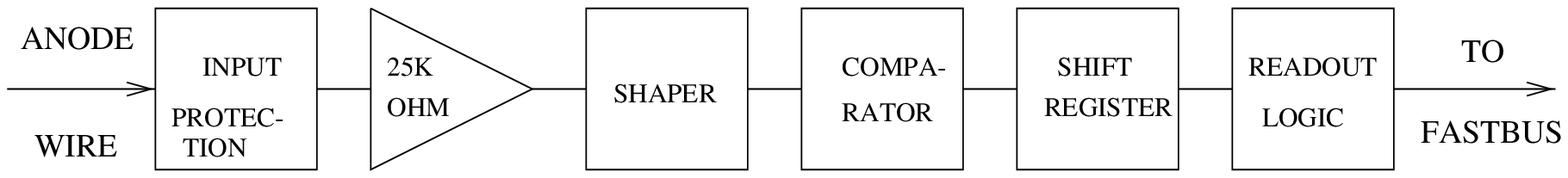,height=3.0in}
	}}
	\caption{Scheme of individual electronics channel.}
	\label{CHANNEL}
	\end{figure}
Beam intensity limitations were imposed by positive ion buildup in the
drift region and in the amplification region of the TPC. (\cite{c,d}). For
the ions created in the amplification region, the gating grid was used to
prevent the transfer of drifting electrons from the drift region to the
amplification region and that of positive ions in the reverse direction
except for selected events. Gating was achieved by applying different
potentials to adjacent wires, as shown in Figure \ref{ENDCAP} collecting
the electrons on the more positive wires. When the trigger system indicates
that an event of interest occurred the gate wires were shorted together
for $40~ms$, a little longer than the maximum drift time of electrons (the
drift velocity of $e^-$ is $1.6~cm/ms$), allowing ionization from the tracks to
reach the amplification region. Because of their slower drift velocity,
positive ions from the amplification at the anode do not reach the gating
grid before the gate is closed, further improving the gating of the TPC.

The electronics were mounted directly on the chamber. The circuit for each
channel, containing an amplifier-shaper-comparator and memory, is shown in
Figure \ref{CHANNEL}. Sixteen channels were assembled in a hybrid package
(LeCroy HTD 161S,M) with inputs spaced $2.54~mm$ apart. Shaping time
constants were selected to match the anode waveform. An integration time
constant of $50~ns$ and a differentiation time constant of $200~ns$ were chosen
in order to minimize noise and maximize two track separation. Each hybrid
contains memory capable of recording 1024 time samples from each channel.
To reduce local noise pickup, all logic levels are ECL. The
clock and address lines are terminated differential ECL. The on-chamber
performance of the hybrid allows input sensitivity of less than $1~\mu A$.   
One MASTER and seven SLAVE hybrid
circuits were assembled on one printed circuit board which is sufficient
for 128 anode readouts and is externally accessed by one clock line, one
data line and 14 control lines. The Master hybrid had  the control logic
necessary to read out itself plus the seven slave hybrids. 
The time samples were obtained at $32~MHz$ and the data is sequentially 
read from each channel's memory at $16~MHz$.
This serial string of data bits is encoded into wire number, drift time and
cluster size by FASTBUS modules in which all rows are encoded
simultaneously. Consequently the entire readout is completed in $10~ms$ (dead
time). More details on the FASTBUS system are given in \cite{e}

The position of the incident beam was measured by two proportional wire
chambers upstream of the MPS magnet. The trigger selected
centrally enriched events for data recording. A plastic Cherenkov counter
$C1-C2$ was used to select Au ions. A halo veto scintillation counter $S2$ was
used to reject events with interactions upstream of the target. The beam
logic was $BEAM = S1\cdot \overline{S2}\cdot C1\cdot S3$. 
In order to enrich the trigger for
central events, a multiplicity counter $S4$ was used to select events in
which the beam particle interacted with the target and the counter $S6$
vetoed fragments with $Z>6$. 
$S4$ was a $10\times 10~m$ scintillation paddle placed just above the beam.
The logic for a trigger was 
$TRIGGER=BEAM\cdot S4\cdot \overline{S6}$. The pulse height in $S6$ was recorded for every event so that tighter
cuts can be applied in the software. Multiplicites of reconstructed tracks
often exceed 200, with an estimated reconstruction efficiency $>90\%$ for long
tracks.

The drift chambers downstream of the TPC's were used for track calibration
of the TPC's. A target was placed in the beam line upstream of the
apparatus to generate a flux of fast parallel tracks filling the TPC
volume. These tracks were measured in the TPC's and drift chambers for
field on and off conditions. Extrapolating the drift chamber tracks to the
TPC hits under these conditions were used to determine the alignment of the
TPC's (field off) and the $\vec E\times \vec B$ corrections (field on) 
due to electric field inhomogeneities at the edges of the TPC's.
\subsection{Data analysis}
	The data and analysis of this experiment were published previously 
	in \cite{e891}. Events with multiplicities of charged tracks
	ranging up to 280 were reconstructed, using a three-dimensional 
	tracking program.
	The total charged track multiplicity at the interaction
	vertex as reconstructed by the tracking program was used as a measure of
	centrality. We found from Monte-Carlo studies that this correlates
	reasonably well with the impact parameter. For the results 
	shown in \cite{e891} we selected events with total charged track
	multiplicities in the acceptance of $>$220,
	corresponding to a measured cross section of
	$270~mb$ (or $\sim 4.4\%$ of the geometric $Au+Au$ cross section).
	This selection criterion resulted in a sample of 14,114 central events.
	
	Tracks which missed the interaction point by more than $7~mm$ were 
	taken to be candidates for a decay vertex. 
	To reduce combinatorial background, we 
	required the decay point to be more than $15~cm$ downstream of the interaction 
	point, the reconstructed momentum vector to point back to the 
	production point within $3.5~mm$ (3 $\sigma$ cut), and the decay plane to 
	be out of
	the magnetic field bend plane. 
	In addition, in order to select tracks which 
	had good momentum resolution, we required the sagitta of the measured tracks 
	to be $>0.67~cm$. The data was recorded with a beam rate of 
	$\sim 10^3 Au~ions/sec$ ($\sim 6\times 10^6$ minimum ionizing). 
	This beam rate resulted in noticeable space charge distortions in 
	the TPC modules in the beam region \cite{c,d}, 
	requiring us to remove from the data
	sample $\Lambda$ hyperons
	decaying with the proton in this region. Specifically, the $\Lambda$ hyperon
	candidates with protons having an azimuthal angle within the range,
	$-52^0<\varphi <149^0$, were removed, where the azimuthal angle, $\varphi$, is 
	measured with respect to $y$ axis.
	All cuts were compensated for in the
	acceptance calculations. 
	The effective masses for $\Lambda$ hyperons were 
	calculated by kinematic hypothesis only, by assigning a proton 
	mass to the positive track and a pion mass to the negative track.

\begin{figure}[h]
	\centerline{\hbox{
	\psfig{figure=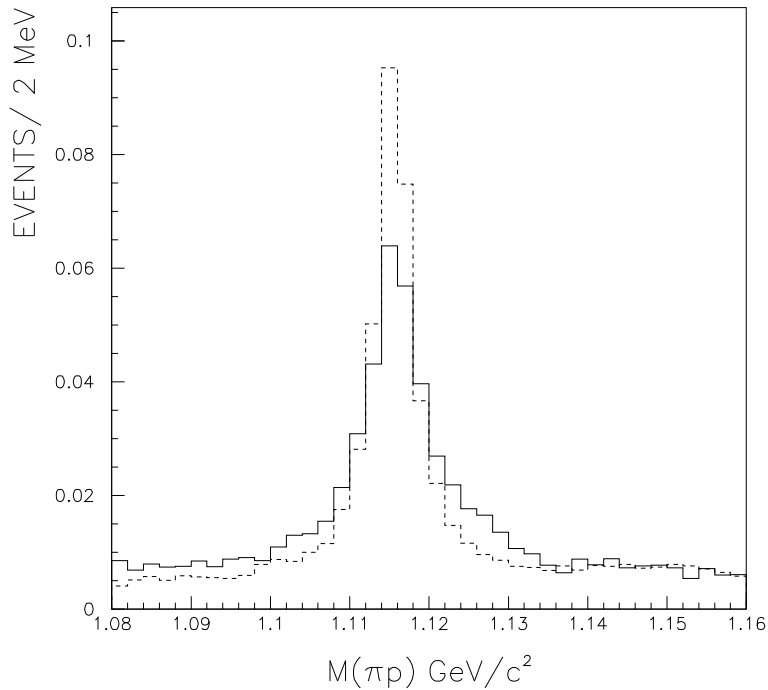,height=4.5in}
	}}
	\caption{Effective mass plot of the proton $\pi^-$ hypothesis for
	$1.4<y(\Lambda)<3.2$ and $m_t-m_0<0.7~GeV/c^2$.
	Data is shown by solid histogram, Monte-Carlo simulation, by dashed line.
	Both curves are normalized per one central event.}
	\label{LAMASS}
	\end{figure}

	Figure \ref{LAMASS} shows the result of the effective mass 
	calculation of the $\Lambda \to p+ \pi^-$ 
	hypothesis for the final selected data sample from the $Au$ target.
	The dashed histogram represents the $\Lambda$ effective mass reconstruction
	from the Monte-Carlo simulation of the experiment.
	The normalization per one central event is done for both data and Monte-Carlo.
	The small discrepancy in the tails is due to uncorrected distortions
	of tracks by the $Au$ beam. The contribution of this discrepancy is
	well within the systematic errors. Different values in the peak of the
	distribution are due to different shapes of the momentum distribution
	in the data and in the model.
	Decay vertices with effective masses in the range of 1.104-1.128 
	GeV/c$^2$ were selected as $\Lambda$ hyperons. 
	The tails of the effective mass
	distributions were used for background subtractions. 
	With the above mentioned cuts we obtained 3932 $\Lambda$ hyperons above
	background for $1.4<y<3.2$ and $m_t-m_0<0.7~GeV/c^2$.
	The $\Lambda$ sample contains a contribution from higher mass states
	which decay to a $\Lambda$ 
	that passes the cuts.
	The main source of these is $\Sigma^0$ decay. This contribution is
	also contained in the model predictions.

	The acceptance corrections of the 
	distributions were calculated using a complete 
	Monte-Carlo simulation of the effects of the apparatus and cuts on 
	the final data sample. Events were generated using a modified HIJET model 
	\cite{shor}. The $(y,m_t)$ shape of the $\Lambda$ spectrum in this model 
	agrees with the measurements of experiment E810 
	\cite{e810} 
	and the strangeness yield is normalized to $K^+$ and $K^-$ measurements
	in $Au+Au$ interactions \cite{e802}. The
	GEANT3 program was used to track the generated tracks through a magnetic field
	and the simulated signals in the TPC were written out. 
	The generated hits included all the known effects of detector
	apertures, efficiencies and distortions.
	(However, see section \ref{eff} below). 
	The results of this simulation were
	then analyzed by the same program used to analyze the actual data, including
	the tracking. A total of 25,000 central interactions were generated and 
	analyzed. The acceptance corrections on the final data sample were done as a 
	function of rapidity and transverse mass. The $\Lambda$ acceptance
	was in the range of 
	2--3\% including a correction for neutral decays.
\subsection{Correction of Track Reconstruction Efficiencies in 
	E891}\label{eff}
	In the original analysis described above \cite{e891} the $\Lambda$ hyperon
	candidates with protons having an azimuthal angle within the range,
	$-52^0<\varphi <149^0$, were removed, where the azimuthal angle, $\varphi$, is 
	measured with respect to $y$ axis and the positive $x$ direction corresponds to
	$\varphi=90^0$. The necessity of this angular cut
	was due to the fact that we could not simulate the dynamic component
	of the $\vec E$ and $\vec E \times \vec B$ distortions 
	caused by the positive ion cloud.
The positive ion cloud charge density evolves on the time scale comparable
to the duration of a spill (300 ms versus 1.2 s). The trigger time is
randomly distributed relative to the beginning of the spill and the number
of beam particles per spill fluctuates. This could not be successfully
simulated in the Monte-Carlo. The angular cuts applied in the original
publication did not guarantee that all the proton tracks will completely
miss the distortion region. For these protons the Monte-Carlo tracks may be
reconstructed with higher efficiency than the data. Since there is a
correlation between the momentum of the $\Lambda$ and 
the track coordinates in the
TPC, this efficiency miscalibration can lead to a change of the measured 
$\Lambda$
spectrum shape, which is crucial for determining the flow parameters. Using
more stringent cuts to eliminate these tracks leads to a data set too
limited to determine the flow parameters.

We have therefore recalculated the efficiency in the region of the
distortion using a  correction based on data from E866 \cite{e866,e866_qm95}. 
The current analysis was also extended to lower rapidity 
than in \cite{e891}. 
	\begin{figure}[h]
	\centerline{\hbox{
	\psfig{figure=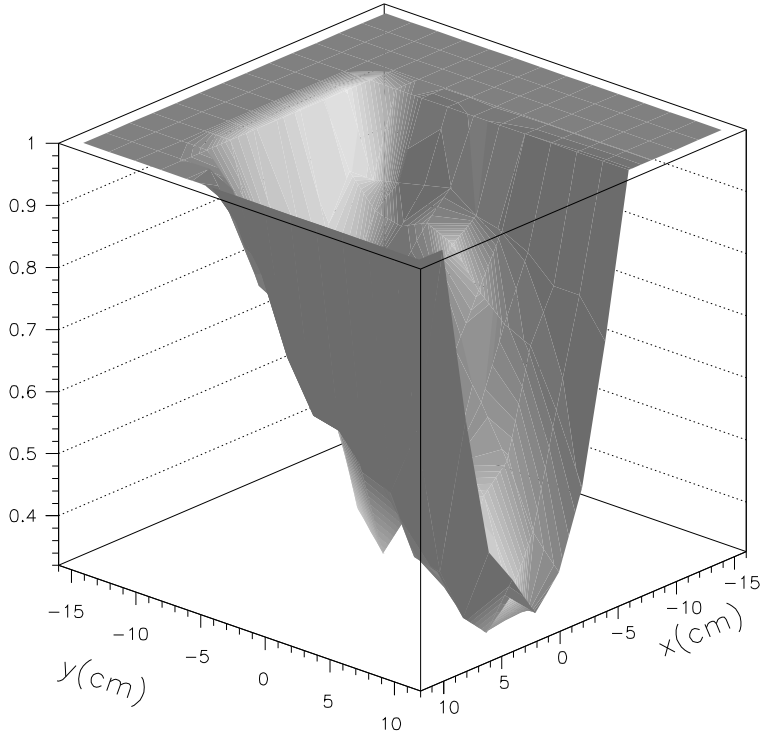,height=4.5in}
	}}
	\caption{Positive track reconstruction efficiency as a function
	of the $x$ and $y$ coordinates of the track intercept with the front
	plane of the TPC. Beam impact coordinates are $x=3.1~cm,~y=0.0~cm$.}
	\label{TRACK_EFF}
	\end{figure}
	\begin{figure}[h]
	\centerline{\hbox{
	\psfig{figure=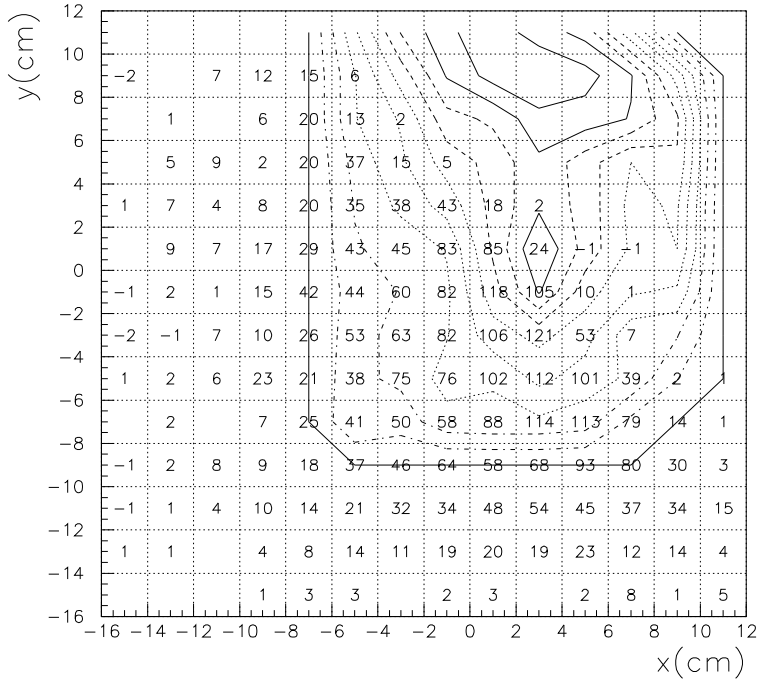,height=4.5in}
	}}
	\caption{Positive track reconstruction efficiency as a function
	of the $x$ and $y$ coordinates of the track intercept with the front
	plane of the TPC. The numbers in each square refer to the number of
	protons from $\Lambda$ decay passing trhough that area. The contours
	represent the constant levels of efficiency ranging from
	1.0, (outermost), to 0.4, (innermost), in steps of 0.075.}
	\label{LAM_MAP}
	\end{figure}
	\begin{figure}[h]
	\centerline{\hbox{
	\psfig{figure=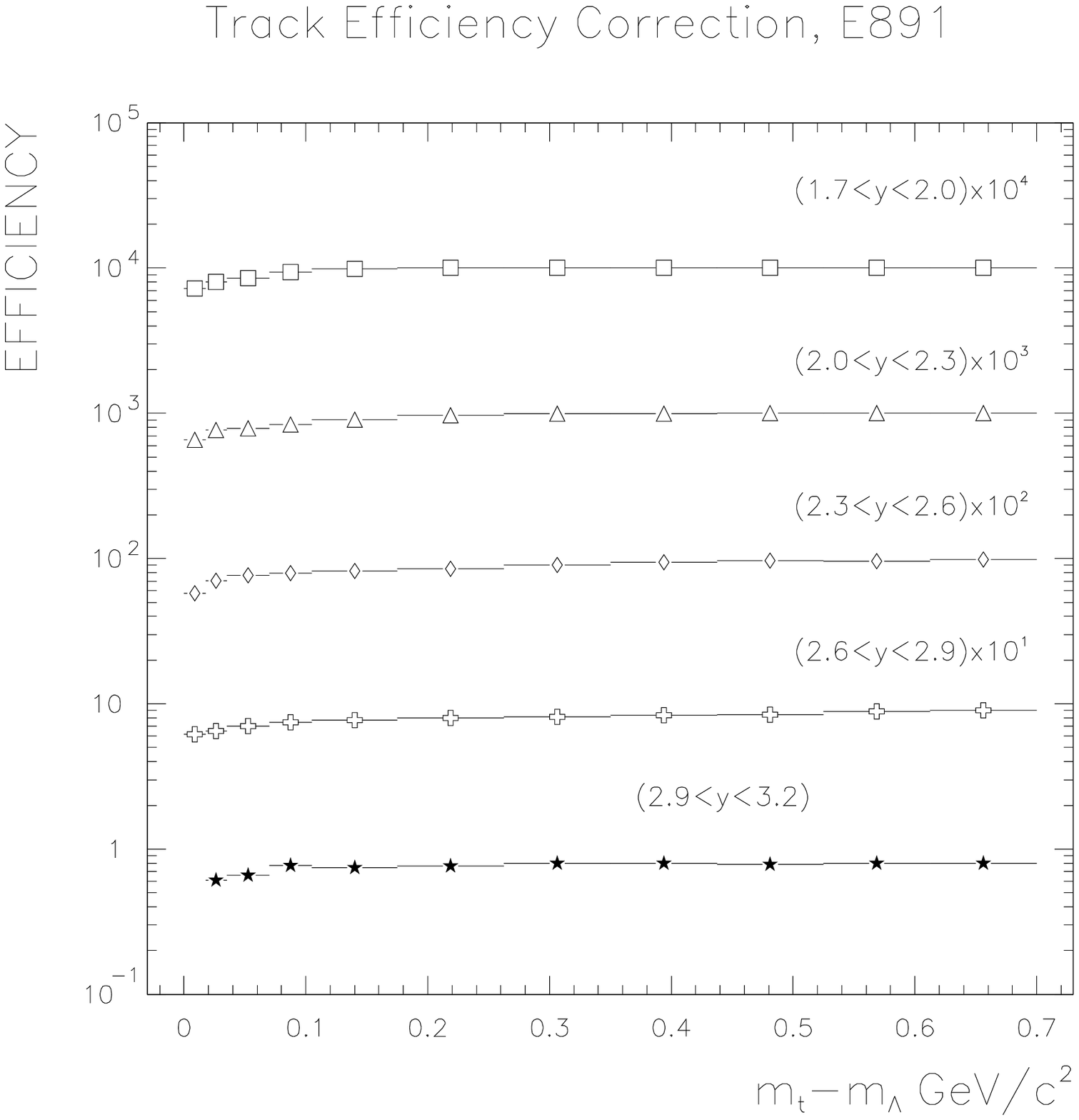,height=4.5in}
	}}
	\caption{$\Lambda$ detection efficiency correction as a function of 
	transverse kinetic energy and rapidity.}
	\label{LAMT_EFF}
	\end{figure}
	The invariant differential spectra for $p$ and $\pi^+$  were
	measured in a wide range of rapidities and transverse momenta by experiment
	E866 and are found to be in a fairly good agreement with the 
	model predictions by "A Relativistic Cascade Model" (ARC) \cite{arc}. 
	We use the 
	ARC model output for the positive particle spectra in order to determine 
	the positive track reconstruction efficiencies as a function of the $x$ 
	and $y$
	coordinates of the track intercept with the front plane of the TPC:
\begin{equation}
\label{eff_track}
E_{tracks}(i,j)=\frac
{N^{tracks}_{DATA}(i,j)/N^{events}_{DATA}}
{N^{tracks}_{{MC}}(i,j)/N^{events}_{MC}},
\end{equation}
	where $N^{tracks}_{DATA}$ is number of positive tracks in the real data,
	$N^{tracks}_{{MC}}$ is that of the Monte--Carlo with ARC used as 
	input and the numbers $(i,j)$ indicate the position of the $2~cm\times 2~cm$ square 
	to be calibrated. The square size of $2~cm$ is sufficiently small
	to insure a continuous dependence of efficiencies as a function of 
	$y$ and $m_t$.
	We perform the cut on the impact parameter, $b_{imp}<3~fm$ on
	the ARC event sample, to achieve the best description of our centrality
	selection criterion and to match the measured interaction cross section.
	Figure \ref{TRACK_EFF} shows the results of the efficiency 
	calculation according
	to (\ref{eff_track}). One can see  significant deterioration 
	of the positive track reconstruction efficiency in the neighborhood 
	of the positive ion sheet. The region plotted was chosen to best display
	the effect in the three dimensional plot. The fact that the efficiency
	goes smoothly to $100\%$ away from the positive ion sheet demonstrates
	that the ARC model successfully extrapolated the E866 data in $y$ and
	$m_t$.
	
	From kinematics it follows that, for relativistic $\Lambda$ hyperons, 
	the momentum of a proton is almost collinear to the momentum of 
	$\Lambda$ which produced it. Thus the incident 
	angle of protons from $\Lambda$'s at the front plane of the 
	first TPC is very close to that of a positive particle which 
	came from the target. 
	To illustrate the considerable overlap of the region of reduced efficiency
	of the positive track reconstruction with the area corresponding to the 
	protons from $\Lambda$ decay, we show the 
	$\Lambda$ counts (background subtracted) as a function of $x,y$
	coordinates in the front plane of the TPC in Figure \ref{LAM_MAP}. 
	A significant fraction of the total number of $\Lambda$ hyperons 
	comes from protons passing through the region of reduced efficiency
	as can be seen from this figure. In the region of ($x>-4~cm$) and
	($y>-6~cm$) the track efficiency drops below 80\%. The number of 
	$\Lambda$'s going through this region, which have passed all cuts
	including the azimuthal angle cut, is about 1376.

\begin{figure}[t]
	\centerline{\hbox{
	\psfig{figure=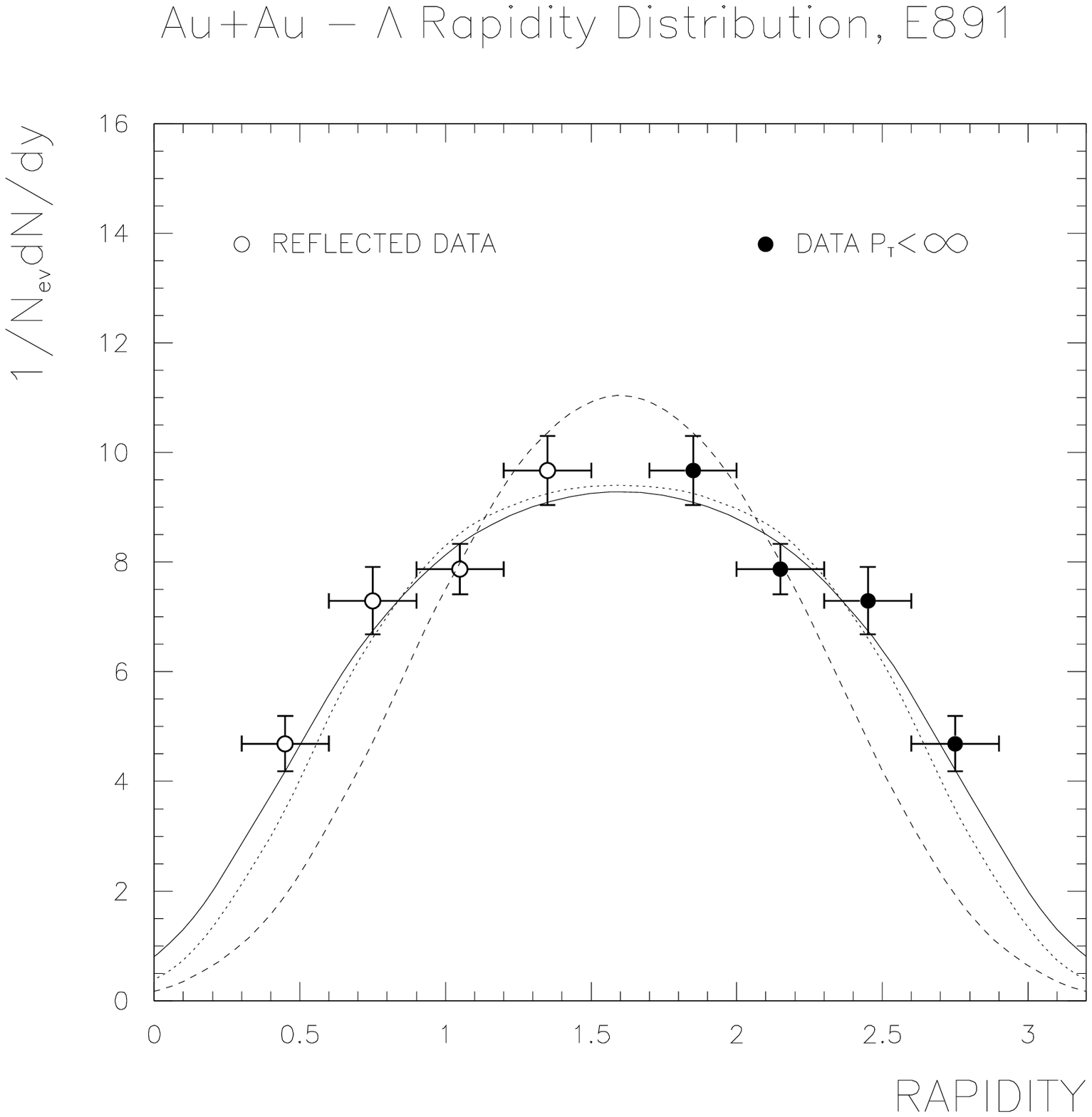,height=4.5in}
	}}
	\caption{Rapidity distribution for the $\Lambda$ hyperon production
	in central $Au+Au$ interactions measured by experiment E891. 
	The models shown are a) RQMD cascade, by dashed lines, 
	b) RQMD with mean fields, by dotted lines 
	c) Blast Wave fit, by solid lines.}
	\label{LAMRAP}
	\end{figure}

	The calculation shows that $\Lambda$'s affected by the inefficiency
	arising from the ion cloud
	have large rapidities and/or low transverse 
	momenta (see Figure \ref{LAMT_EFF}). 
	The effective values of efficiency corrections for rapidity bins,
	integrated over $m_t$, 
	$(1.7<y<2.0)$, $(2.0<y<2.3)$, $(2.3<y<2.6)$ and $(2.6<y<2.9)$ 
	are 1.03, 1.10, 1.21 and 1.36 respectively.
	The systematic errors for the $\Lambda$ spectra, defined as the 
	difference between the efficiency corrected data and the data 
	without efficiency correction
	\cite{e891}, do not exceed the statistical ones. 
	The double differential spectra and rapidity distribution corrected for
	the track reconstruction efficiency are shown in Figures \ref{LAMT} and
	\ref{LAMRAP}.
\section{The Blast Wave Parameterization}
	An interesting phenomenon observed in relativistic $Au+Au$ collision 
	by experiment E891 is low-$m_t$ suppression in the double differential 
	spectra of $\Lambda$ hyperons around midrapidity, a deviation from a single
	exponential scaling valid for lighter nuclei \cite{e810} 
	and for $pp$ interactions. The analogous effect has been observed
	by experiment E866 for protons \cite{e866}. It was suggested 
	in \cite{e866} that the effect is due to the strong transverse
	expansion. 
	The influence of collective nuclear flow has been established at
	different accelerator facilities. For instance, for the explanation 
	of the different inverse slopes
	in the momentum distributions for protons and pions at Bevalac 
	in the central $Ne+NaF$ collisions at the lab energy of $800~MeV$ 
	per nucleon, Siemens and Rasmussen \cite{siemens} suggested that 
	after thermal equilibration the nuclear matter experiences 
	spherically symmetric expansion. 
	According to the Bjorken scenario of one dimensional longitudinal 
	expansion \cite{bjorken}, the spectra of particles produced in 
	ultra-relativistic heavy-ion collisions are boost invariant. 
	Such a spectrum can be represented as superposition of the 
	thermal sources distributed uniformly over a limited interval 
	of the longitudinal boost angle, $\eta$ \cite{stachel}
\footnote{The longitudinal boost angle, 
	$\eta =0.5\ln{[(1+v_z)/(1-v_z)]}$, is essentially the rapidity 
	of the moving source.}:
\begin{equation}
\label{long_boost}
{\cal E}\frac{d^3N}{dp^3}=
\int\limits_{-\eta_{max}}^{\eta_{max}}
d\eta m_t\cosh{(y-\eta)}e^{-m_t\cosh{(y-\eta)}/T},
\label{long}
\end{equation}
	where rapidities $y$ and $\eta$ are measured in the center of mass frame.
	The limits of the boost invariance interval, $[-\eta_{max},\eta_{max}]$,
	are confined between the projectile and the target rapidities. 
	Assuming that the
	longitudinal and the transverse expansion can be decoupled, P.
	Braun-Munzinger et al \cite{stachel} used expression
	(\ref{long}) integrated over the transverse mass squared 
	to evaluate the rapidity distributions of different hadrons
	in $Si+Al$ interactions at AGS energy. 
	Fairly good description of the rapidity 
	distributions was achieved with the temperature,
	$T=120~MeV$, and  the mean longitudinal expansion velocity, $<v_l>=0.52$.
	However, no attempt to parameterize the double
	differential distributions in a wide range of rapidities was made in this
	article. For the description of invariant differential multiplicity at 
	midrapidity the authors applied expression (A8) from \cite{heinz}, 
	which is intended for the description of the transverse mass distribution
	integrated over rapidity, not the invariant differential distribution!
	The disadvantage of such parameterization is that sources at different
	rapidities actually interfere with each other. 	
	Unlike the SPS energies, where a modest boost of an observer in the 
	longitudinal direction
	does not change the velocities of the receding nucleon pancakes, 
	the AGS energy is not sufficiently high 
	for the boost invariant expansion to take place. 
	In addition, nonzero  curvatures of the reduced invariant differential
	multiplicities of baryons, 
	${\cal E}/m_t\cdot d^3N/dp^3$, on a logarithmic scale,
	indicate the presence of a transverse radial expansion with the strength
	varying as a function of rapidity. 
	The longitudinal and the transverse expansion may be decoupled 
	accurately only in the limit of nonrelativistic transverse velocities.
	Therefore, we believe that the correct procedure of determining the flow 
	velocity profile is a parameterization of the double differential spectra 
	of particles. In this procedure the rapidity distributions and integrated
	transverse mass distributions will be described automatically.

	Usually the transverse velocity is extracted by
	fitting the spectrum assuming a power law of the transverse expansion
	velocity $v_t\propto r_t^{\alpha}$ and a constant density, $\rho(r_t)$
	(see, for example \cite{stachel}). 
	From the cascade simulations it is known that the density
	resembles a gaussian profile, $\rho(r_t)=\rho_0e^{-r_t^2/2\sigma^2}$, 
	which can lead to significant consequences to
	the parameters obtained from a fit \cite{qmd}. For the traditionally used 
	uniform density profile, as well as for a gaussian density distribution,
	the integration over the 
	transverse expansion velocity selects non-zero velocities due to the 
	rising factor, $2\pi v_t$. The probability for a particle to have 
	a transverse collective velocity, $v_t$, has a peak at a value defined
	by parameters $\alpha$ and $\sigma$ in case of a gaussian density.
	Our data on the double differential $\Lambda$ multiplicities, however
	indicates that a stronger selection of the transverse expansion velocities
	is present in reality, which is consistent with the creation of a transverse blast wave
	for the central $Au+Au$ interactions. The development of the blast wave 
	due to the sudden creation of hot dense matter 
	was originally advocated in \cite{siemens} where the spherical blast wave 
	scenario was used to account for different apparent temperatures for 
	$p$ and $\pi$. While apparent temperatures, the asymptotic values of 
	the inverse slopes of reduced invariant differential cross sections,
	are not strongly sensitive to the probability distribution for collective
	velocities, the low-$m_t$ slopes are.  
	It was pointed out in \cite{matiello} that the step-like freeze-out
	distance distribution, widely used for experimental parameterization 
	of transverse expansion (see for example \cite{stachel}) 
	is quite different from the one observed in RQMD,
	where the freeze-out close to the collision axis is suppressed. 
	Before experiencing the final interaction the particles are propagated to 
	the periphery of the system, where they acquire  
	the essential collective motion \cite{matiello}.
	This statement, together with the power-like transverse velocity profile is
	consistent with a blast wave scenario. It was also claimed in \cite{matiello}
	that it is impossible to describe a spectrum with a single temperature
	and transverse expansion velocity. We will demonstrate that the spectrum
	of $\Lambda$ hyperons can be described in terms of an aspherical blast wave
	with a single temperature.

	The influence of the transverse blast wave on the low-$m_t$ slopes 
	of the reduced invariant differential cross sections can be understood by 
	noting that the spectrum in a given rapidity bin originates primarily from sources,
	macroscopic elements of nuclear matter, with a limited  
	rapidity range. One can thus follow this influence by considering the
	transversely expanding ring at this given rapidity.
	Boosting the invariant Boltzmann distribution in the transverse direction 
	$\varphi$ by velocity $v$,
\begin{equation}
{\cal E}\frac{d^3N_{\varphi}}{dp^3}=\gamma ({\cal E}-p_t v \cos{\varphi})
e^{-\gamma ({\cal E}-p_t v \cos{\varphi})/T},
\end{equation}
	and integrating this expression over the azimuthal angle, 
	$\varphi$, one can derive the equation for a spectrum from 
	the transversely expanding ring:
\begin{equation}
{\cal E}\frac{d^3N}{dp^3}=2\pi\gamma 
\left\{
{\cal E} I_0(\xi)-p_tvI_1(\xi)
\right\}e^{-\frac{\gamma {\cal E}}{T}},
\label{trans}
\end{equation}
where 
\begin{equation}
\xi=\gamma p_tv/T,~~~\gamma=1/\sqrt{1-v^2},
\end{equation}
and functions 
$I_0(x)=(2\pi)^{-1}\int\limits_{0}^{2\pi}e^{x\cos{\varphi}}d\varphi$ and 
$I_1(x)=(2\pi)^{-1}\int\limits_{0}^{2\pi}\cos{\varphi}e^{x\cos{\varphi}}d\varphi$ 
	are the modified Bessel functions of the zeroth and the first order respectively.
	The low-$m_t$ slopes for the of reduced invariant differential 
	cross section can be found by resolving the logarithm of equation
	(\ref{trans}) divided by $m_t$ 
	into a Taylor series in the vicinity of $m_t\to m_0$, 
	assuming ${\cal E}=m_t$, and saving only the linear term:
\begin{equation}
\ln\left(\frac{\cal E}{m_t}\frac{d^3N}{dp^3}\right)_{m_t\to m_0}=
const+\frac{\gamma \{\gamma m_0 v^2  - 2 T (1 + v^2)\}}
{2T^2}(m_t-m_0).
\end{equation}
	\begin{figure}[h]
	\centerline{\hbox{
	\psfig{figure=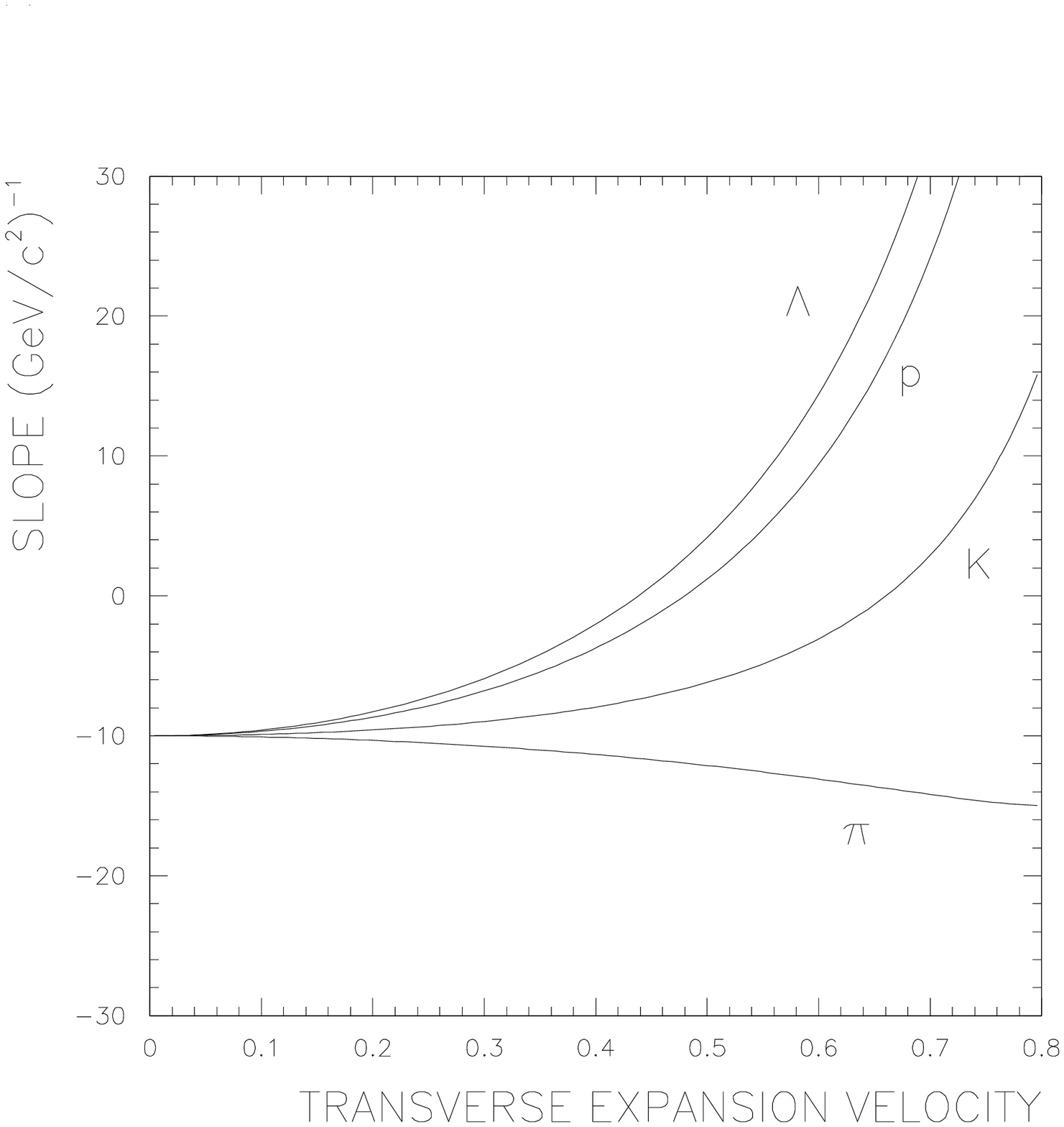,height=4.5in}
	}}
	\caption{The low-$m_t$ slopes for various hadrons as a function
	of the transverse expansion velocity for a fixed temperature of $T=100~MeV$.}
	\label{SLOPE}
	\end{figure}
	The factor multiplying the transverse kinetic energy, $m_t-m_0$, 
	represents the low-$m_t$ slope of the reduced invariant 
	differential cross section on a logarithmic scale.
	On Figure \ref{SLOPE} the low-$m_t$ slopes are shown for various hadrons 
	as a function of the transverse expansion velocity for a fixed temperature 
	of $T=100~MeV$. One can see that for sufficiently high transverse expansion 
	velocities the slopes become positive, 
	yielding the low-$m_t$ suppression. 
	It is worth noting that for $\pi$ mesons the low-$m_t$
	slope shows the opposite behavior, though of smaller magnitude. 
	Generally, the condition $m_0>2T$ must be satisfied
	for low-$m_t$ suppression to take place. The higher the temperature of
	the hadron gas the less affected is the low-$m_t$ slope of the spectrum.

\begin{figure}[t]
	\centerline{\hbox{
	\psfig{figure=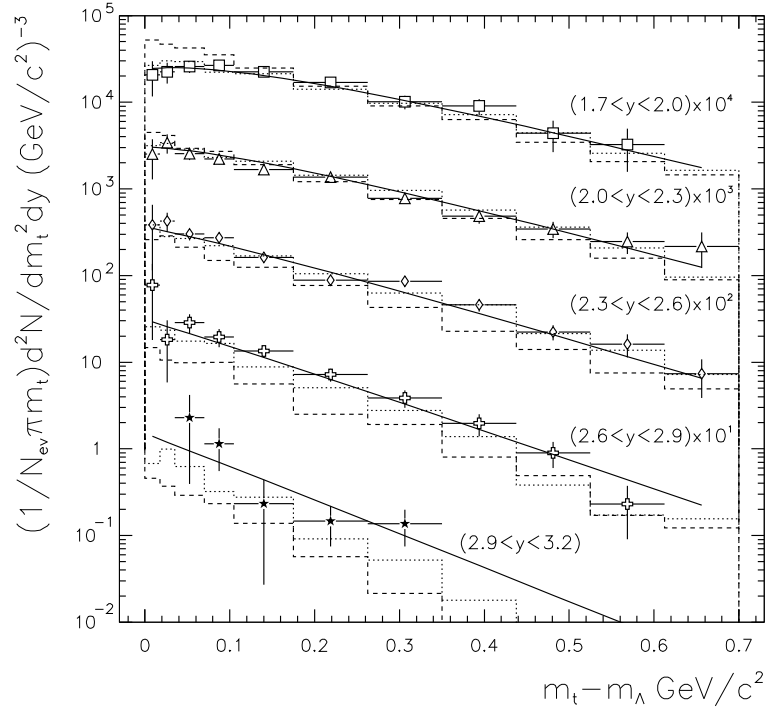,height=4.5in}
	}}
	\caption{Double differential multiplicities for $\Lambda$ production
	in central $Au+Au$ interactions measured by experiment E891. 
	The models shown are a) RQMD cascade, by dashed lines, 
	b) RQMD with mean fields, by dotted lines 
	c) Blast Wave fit, by solid lines.}
	\label{LAMT}
	\end{figure}

	We represent the inclusive spectrum as 
	a continuous superposition of flowing Boltzmann sources with 
	an effective temperature, $T$. 
	The effective flow is described by an azimuthally symmetric 
	surface in the space of the longitudinal boost angle, $\eta$,
	and the transverse two-dimensional velocity, the Blast Wave. 
	The blast wave in our approach belongs to the 
	interval, $[-\eta_{max},\eta_{max}]$, where both limits are measured in  
	$Au+Au$ center of mass system. An expression for the invariant
	differential multiplicity reads as follows:
\begin{equation}
\label{3d_expansion}
{\cal E}\frac{d^3N}{dp^3}=
\int\limits_{-\eta_{max}}^{\eta_{max}}
d\eta {\cal E}f(y,\eta,m_t),
\label{smoke}
\end{equation}

\begin{equation}
{\cal E}f(y,\eta,m_t)=\frac{1}{\pi}A(\eta)
\{\chi I_0(\xi)-\xi I_1(\xi)\}
e^{{\frac{m_0}{T}-\chi}},
\label{single_smoke}
\end{equation}
\begin{equation}
\chi=\frac{\gamma_t(\eta)~m_t~\cosh{(y-\eta)}}{T},~~
\xi=\frac{\gamma_t(\eta)~p_t~v_t(\eta)}{T},
\end{equation}
\begin{equation}
\label{3d_last}
\gamma_t(\eta)=\frac{1}{\sqrt{1-v_t^2(\eta)}}.
\end{equation}
	Choosing the flow velocity profile and the rapidity 
	distribution of the emission power of the blast wave,
	we employ the following simple functions: 
\begin{equation}
v_t(\eta)=v_0(1-\eta^2/\eta_{max}^2)^{\nu},~~
A(\eta)=A_0(1-\eta^2/\eta_{max}^2)^{\alpha},
\end{equation}
	where $\alpha,~\nu \ge 0$. Rapidities, $y$ and $\eta$, are 
	measured in the center of mass frame of the collision system and 
	the transverse velocity, $v_t$, is measured 
	in a system co-moving with $\eta$. 
	The functions $v_t(\eta)$ and $A(\eta)$ 
	are chosen to allow one to tune the shape of the blast wave to 
	the data. One can easily see that with $\nu=1/2$
	and $v_0=\eta_{max}$
	the function $v_t(\eta)$ describes a spherical blast wave, where for 
	$\alpha=0$ the emission power is uniformly distributed along the surface
	of the blast wave. With $\eta_{max}\to 0$ and $v_0>0$, 
	one encounters purely transverse expansion, as in the Landau scenario
	\cite{landau}. 
	It can be also seen that the boost invariant Bjorken cylinder 
	can be described with $\nu=\alpha=0$.  
	With positive values of parameters $\nu$ and $\alpha$, 
	the functions above reflect the experimental fact of increase
	of the transverse expansion velocity towards midrapidity, which is
	seen from the increase of the low-$m_t$ slope of the reduced invariant 
	differential cross section on a logarithmic plot (see Figure \ref{LAMT}).

	Mean rapidity and the transverse velocity in the blast wave 
	are given by the following expressions:
\begin{eqnarray} 
\label{mean_rap}
<\eta>=&\frac{\int\limits_{0}^{1}x(1-x^2)^{\alpha}dx}
{\int\limits_{0}^{1}(1-x^2)^{\alpha}dx}\eta_{max}=&
\frac{\Gamma(\alpha+3/2)}{\sqrt{\pi}\Gamma(\alpha+2)}
\eta_{max}, \\
\label{mean_vt}
<v_t>=&\frac{\int\limits_{0}^{1}(1-x^2)^{\alpha+\nu}dx}
{\int\limits_{0}^{1}(1-x^2)^{\alpha}dx}v_0=&
\frac{\Gamma(\alpha+\nu+1)}{\Gamma(\alpha+\nu+3/2)}
\frac{\Gamma(\alpha+3/2)}{\Gamma(\alpha+1)}v_0,
\end{eqnarray}
	where averaging is performed over the forward hemisphere in the center of
	mass system.
	The rapidity distribution and the 
	total number of particles emitted by this source are
\begin{equation}
\label{3d_rap}
\frac{dN(y)}{dy}\approx
\frac{m_0^2e^{\frac{m_0}{T}}}{\pi}
\int\limits_{-\eta_{max}}^{\eta_{max}}d\eta
A(\eta)\frac{dN_{boltz}(y-\eta)}{dy},
\end{equation}
\begin{equation}
\label{n_total}
N= \frac{A_0m_0^2\eta_{max}e^{\frac{m_0}{T}}}{\sqrt{\pi}}
\frac{\Gamma(\alpha+1)}
     {\Gamma(\alpha+3/2)}
N_{boltz},
\end{equation}
	respectively, 
	where $\Gamma(x)=\int\limits_{0}^{\infty}t^{x-1}e^{-t}dt$, $(x>1)$, 
	is the gamma function, and
\begin{equation}
\frac{dN_{boltz}(y)}{dy}=
2 \pi \left(1+\frac{2\cdot T}{m_0\cosh{(y)}}+
\frac{2\cdot T^2}{m_0^2\cosh^2{(y)}}\right)
e^{-\frac{m_0\cosh(y)}{T}}
\end{equation}	
is the Boltzmann rapidity distribution.

	We have applied the formulas listed in this paragraph to obtain the 
	expansion parameters as well as the mean temperature of $\Lambda$ hyperons
	after the freeze-out. The obtained parameters
	$A_0$, $v_0$, $\alpha$, $\nu$, $T$, $\eta_{max}$ are listed in Table 
	\ref{BLAST} together with the mean source rapidity, $<\eta>$, 
	the mean transverse expansion velocity, $<v_t>$, and 
	the integrated yield, $N_{\Lambda}$ derived from 
	the above six parameters.
\begin{table}[t]
\begin{center}
\begin{tabular}{|c|c|c|c|} \hline
       	          &  E891           &      RQMD cascade  & RQMD mean fields  \\ \hline \hline
 $A_0~(GeV/c^2)^{-2}$ & $4.42\pm 1.0           $& $4.40\pm 0.76$ & $4.36 \pm 0.52$ \\ \hline
 $v_0        	  $ & $0.50\pm 0.08          $& $0.34\pm 0.06$ & $0.49 \pm 0.03$ \\ \hline
 $\alpha        	  $ & $0.50\pm ^{1.17}_{0.50} $& $0.98\pm^{1.4}_{0.98}$ & $0.22 \pm^{0.31}_{0.22}$ \\ \hline
 $\nu        	  $ & $0.75\pm ^{1.51}_{0.75}$& $0.98\pm^{1.48}_{0.98}$ & $0.57 \pm 0.22$ \\ \hline
 $T~(MeV/c^2)       $ & $96  \pm 37            $& $136  \pm 33 $ & $95   \pm 13   $ \\ \hline
 $\eta_{max}    	  $ & $1.36\pm 0.38          $& $1.1\pm 0.30 $ & $1.18 \pm 0.10$ \\   \hline \hline
 $\chi^2/NDF 	  $ & $32/43                 $& $58/58       $ & $60/47        $ \\   \hline \hline
 $<v_t> 	 	  $ & $0.40\pm 0.12           $& $0.27\pm0.08 $ & $0.40\pm0.04 $ \\ \hline
 $<\eta> 	 	  $ & $0.58\pm 0.07           $& $0.42\pm0.02 $ & $0.54 \pm0.02 $ \\ \hline
 $N_{\Lambda} 	  $ & $20\pm 3.33               $& $18\pm 1.52   $ & $19   \pm 1.12   $ \\ \hline
\end{tabular}
\centerline{}
\caption{Results of the Blast Wave parameterization of $\Lambda$ double differential
		multiplicities for E891 data and RQMD predictions.
		Six main parameters are listed first. The mean transverse expansion
		velocity, $<v_t>$, longitudinal boost angle, $<\eta>$, and mean
		multiplicity, $N_{\Lambda}$, are obtained using formulas 
		(\ref{mean_rap}), (\ref{mean_vt}) and (\ref{n_total}).}
\label{BLAST}
\end{center}
\end{table}
\section{Results and Discussions}
	In Table \ref{BLAST} the results of the parameterizations of the double 
	differential multiplicities for the $\Lambda$ production measured by E891
	are listed together with the results obtained from the cascade 
	version of RQMD and the version of RQMD with mean field potentials
	turned on. 
	For comparison with RQMD model predictions we have selected the sample
	of central events with the cut on the impact parameter, $b<3~fm$. This 
	value of the cut was chosen, since the resulting $Au+Au$ geometric cross 
	section, $\pi b_{max}^2$, matches the cross section we obtain
	and the resulting $\pi^-$ differential multiplicities agree well 
	with our measurements 
\footnote{In Reference \cite{e891} the cut of $b<4~fm$ was chosen in models for
comparison with our data.}.
	One can see that the cascade version of RQMD underpredicts 
	the longitudinal and the transverse flow, which can be verified 
	by comparing the mean transverse expansion velocity, $<v_t>$, and the 
	mean longitudinal boost angle, $<\eta>$. This fact can also be seen 
	on the plots of double differential multiplicity for $\Lambda$ production
	(Figure \ref{LAMT}) where the cascade version of RQMD fails to describe the low-$m_t$
	suppression effect in the midrapidity region. The rapidity distribution of 
	the $\Lambda$ hyperon measured by E891 (Figure \ref{LAMRAP}) has a larger width than predicted by 
	RQMD cascade, which corresponds to the smaller strength of longitudinal 
	flow in the cascade. One can also see in Table \ref{BLAST} that all the 
	parameters of the Blast Wave obtained by fitting the results of RQMD
	with mean fields turned on agree with the ones obtained by fitting the data
	within experimental uncertainties. The low-$m_t$ suppression effect 
	is reproduced by this version of RQMD and the width of the rapidity
	distribution agrees well with the experimentally measured width.
\section{Conclusions}
	Based on the agreement of the predictions of RQMD with mean fields 
	with our measurements for the double differential and rapidity 
	distributions for $\Lambda$ production in the central $Au+Au$ 
	interactions and underestimation of the flow strengths by RQMD
	cascade we can confirm that the mean field effects play important
	role in the dynamics of the $Au+Au$ interactions at the AGS energy.	
\begin{ack}
	We wish to thank all members of E891 collaboration for use of their 
	raw data. We are grateful to
	H. Sorge for access to the RQMD code and advice on running it. 
	We thank D. Kahana for providing us with events from the ARC model.
	One of us (S.V. Efremov) would like  to thank H. Sorge, D. Rischke, 
	and J. Wessels for the useful discussions, 
	and W. Llope for expressing interest in this work. 
	Thanks are also due to  
	M. Pollack and T. Vongpaseuth for generating 
	and providing to us the RQMD events with mean fields potentials.
\end{ack}

\end{document}